\documentclass[10pt,a4paper]{article}

\begin{document}

\markboth{M. Arik, U. Kayserilioglu}
{The anticommutator spin algebra, representations and invariance}

\title{The anticommutator spin algebra, its representations and quantum group invariance}
\author{M. Arik, U. Kayserilioglu\\
       {\it Bo\~ gazi\c ci University, Physics Department}\\
       {\it Bebek 80815, Istanbul, Turkey} \\}
\date{\today}

\maketitle
\begin{abstract}
We define a 3-generator algebra obtained by replacing the commutators by anticommutators in the
defining relations of the angular momentum algebra. We show that integer spin representations are in
one to one correspondence with those of the angular momentum algebra. The half-integer spin representations,
on the other hand, split into two representations of dimension $j + \frac12$. The anticommutator spin algebra
is invariant under the action of the quantum group $SO_q(3)$ with $q=-1$.
\end{abstract}

\newcommand{\beq}{\begin{equation}}
\newcommand{\eeq}{\end{equation}}
\newcommand{\bea}{\begin{eqnarray}}
\newcommand{\eea}{\end{eqnarray}}
\newcommand{\anti}[2]{\{#1, #2\}}
\newcommand{\ket}[1]{\mid #1\;\rangle}
\newcommand{\bra}[1]{\langle\; #1\mid}
\newcommand{\braket}[2]{\langle\; #1\mid #2\;\rangle}

\section{Introduction}

The algebra of observables in quantum theory plays a fundamental
role. When classical systems are quantized, their classical
symmetry algebra acting on a set of physical observables, in
simplest examples, remains the same. For some
completely integrable non-linear models, consistent quantization
requires that the classical symmetry group be replaced by a
quantum group \cite{frt,drinfeld,woronowicz,manin}
via a deformation parameter $q = 1 + O(\hbar)$. In recent years quantum
groups involving fermions have received widespread attention. These include
deformed fermion algebras \cite{jx,xh,sm,chung}, spin chains
\cite{nt,gppr,bnnpsw} and Fermi gases \cite{ubriaco}. At
the same time, some quantum systems, most notably fermionic
quantum systems do not have any classical analogues. Nevertheless, fermions
are perhaps the most important sector of quantum phenomena. Motivated by
these considerations, we define a fermionic version of the
angular momentum algebra by the relations
\pagebreak
\bea
\anti{J_1}{J_2} & = & J_3 \label{eqn:defrel1} \\
\anti{J_2}{J_3} & = & J_1 \label{eqn:defrel2} \\
\anti{J_3}{J_1} & = & J_2 \label{eqn:defrel3}
\eea
where $J_1$, $J_2$, $J_3$ are hermitian generators of the algebra. We will name this algebra
ACSA, the anticommutator spin algebra. In these expressions the curly bracket denotes the anticommutator
\beq
\anti{A}{B} \equiv AB + BA
\eeq
so (\ref{eqn:defrel1}-\ref{eqn:defrel3}) should be taken as the definition of an associative
algebra. This proposed algebra does not fall into the category of
superalgebras in the sense of Berezin-Kac axioms. In particular, the algebra is consistent without
grading and there are no (graded) Jacobi relations. As it is defined this algebra falls into
the category of a (non-exceptional) Jordan algebra where the Jordan product is defined by:
\beq
A \circ B \equiv \frac12 (AB + BA) \quad .
\eeq
A formal Jordan algebra, in addition to a commutative Jordan product, also satisfies
$A^2\circ(B\circ A) = (A^2\circ B)\circ A$. When the Jordan product is given in terms of an
anticommutator this relation is automatically satisfied.
Just as a Lie algebra where the Lie bracket as defined by the commutator leads to an enveloping
associative algebra, a Jordan algebra defined in terms of the above product leads to an enveloping
associative algebra which we consider as an algebra of observables.

The physical properties of this system turn out to
be similar to those of the angular momentum algebra yet exhibit remarkable differences.
Since the angular momentum algebra is used to describe various internal symmetries, ACSA could
be relevant in describing those symmetries.

In section 2 we will show that ACSA is invariant under the action of the quantum group $SO_q(3)$
with $q=-1$. Here, $SO_q(3)$ is defined as the quantum subgroup of $SU_q(3)$ where each of the
(non-commuting) matrix elements of the $3$x$3$ matrix is hermitian. We note that this defines a
quantum group only for $q=\pm1$. For $q=1$ one has the real orthogonal group $SO(3)$.

In section 3, we will construct all representations of ACSA and show that the representations
can be labelled by a quantum number $j$ corresponding to the eigenvalue of $J_3$ whose absolute
value is maximum. For integer $j$, spectrum of $J_3$ is given by $j, j-1, \ldots, -j$ whereas
for half-integer $j$ there are two representations. These two representations are such that for
$j = 2k\pm\frac12$ spectrum of $J_3$ is respectively
given by $j, j-2, \ldots, \pm\frac12$ and $-j, j+2, \ldots, \mp\frac12$.
Section 4 is reserved for conclusions and discussion.

\section{The invariance quantum group $SO_q(3)$, $q= -1$}

In order to find the invariance quantum group of this algebra, we transform the generators $J_i$ to $J'_i$ by:
\beq \label{trans}
J'_i = \sum_j\alpha_{ij} J_j \quad .
\eeq
The matrix elements $\alpha_{ij}$ are hermitian since $J_i$'s are
hermitian and they commute with $J_i$'s but do not commute with each other.
For the transformed operators to obey the
original relations, there should exist some conditions on the $\alpha$'s which define the invariance quantum group of the
algebra. It is very convenient at this moment to switch to an index notation that encompasses all three defining
relations of the algebra in one index equation. For the angular momentum algebra this is possible by defining the
totally anti-symmetric rank 3 pseudo-tensor $\epsilon_{ijk}$. A similar object for ACSA which we will call the fermionic
Levi-Civita tensor, $u_{ijk}$, is defined as:
\beq
u_{ijk} =
\left\{
\begin{array}{cl}
1, & i \neq j \neq k, \\
0, & \mbox{otherwise.}
\end{array}
\right.
\eeq
Thus the defining relations (\ref{eqn:defrel1}-\ref{eqn:defrel3}) become:
\beq
\anti{J_i}{J_j} = \sum_k  u_{ijk}\,J_k + 2\delta_{ij}\,J_i^2
\eeq
The second term on the right is needed since when $i = j$ the left-hand side becomes $2J_i$.
When we apply the transformation (\ref{trans}) on this relation we get:
\beq
\anti{J'_k}{J'_m} = \sum_p u_{kmp}J'_p = \sum_{p,\; j} u_{kmp}\,\alpha_{pj}\,J_j\quad\quad\mbox{for}\quad k\neq m.
\eeq
However, substituting the transformation equations into the left-hand side, we have:
\beq
\anti{J'_k}{J'_m} = \sum_{i,\;j}\left([\alpha_{ki}, \alpha_{mj}]J_iJ_j + \alpha_{mj}\,\alpha_{ki}\,u_{ijn}\,J_n + 2\alpha_{mj}\,\alpha_{kj}\,J_j^2\right)
\eeq
These two equations yield the following relations among $\alpha_{ij}$ when $k\neq m$:
\bea
\anti{\alpha_{mj}}{\alpha_{kj}} & = & 0 \label{invrel1} \\
\left[\alpha_{ki}, \alpha_{mj}\right]& = & 0 \quad\quad \mbox{for}\quad i \neq j \label{invrel2} \\
\sum_{i\; j}\alpha_{mj}\,\alpha_{ki}\,u_{ijn}  & = & \sum_p u_{mkp}\,\alpha_{pn} \label{invrel3}
\eea

Now we will define the quantum group $SO_q(3)$ and show that the relations above correspond to the case $q=-1$.
The quantum group $SO_q(3)$ can be defined as the quantum subgroup of $SL_q(3, C)$ where an element is given by:
\beq
A =
\left(
\begin{array}{ccc}
\alpha_{11} & \alpha_{12} & \alpha_{13} \\
\alpha_{21} & \alpha_{22} & \alpha_{23} \\
\alpha_{31} & \alpha_{32} & \alpha_{33}
\end{array}
\right)
\eeq
where
\beq
\alpha_{ij}^* = \alpha_{ij}
\eeq
\beq
A^T = A^{-1}
\eeq
and
\beq \label{gl2}
\left(
\begin{array}{cc}
\alpha_{mj} & \alpha_{mi} \\
\alpha_{kj} & \alpha_{ki}
\end{array}
\right)
\in
GL_{q}(2)\quad\mbox{for}\quad k\neq m, i\neq j.
\eeq

The quantum group $SO_q(3)$ is equivalent to the quantum group \break\mbox{$SL_q(3, R) \cap SU_q(3)$}. However
one can show for $SL_q(3)$ that $q= e^{i\beta}$ for some $\beta \in R$ and similarly for $SU_q(3)$ that
$q \in R$. Thus one finds that $q = \pm 1$ for $SO_q(3)$. When $q = 1$ the quantum group becomes
the usual $SO(3)$ group; the interesting case is when $q = -1$ which, as we will show, is the invariance quantum
group of ACSA.

Equations (\ref{invrel1}) and (\ref{invrel2}) are easily shown to be satisfied by the matrix $A \in SO_{q=-1}(3)$ by recognizing that
the quantities involved belong to a submatrix that is an element of $GL_{q=-1}(2)$, as in equation (\ref{gl2}).
For a general matrix $M \in GL_q(2)$ where:
\[
M =
\left(
\begin{array}{cc}
a & b \\
c & d
\end{array}
\right)
\]
we have the relations:
\bea
a c & = & q c a \label{ac} \\
a d - q b c & = & d a - q^{-1} c b \label{det} \\
b c & = & c b \label{bc}
\eea
The relation (\ref{ac}) implies that $\alpha_{mj}\;\alpha_{kj} = (-1) \alpha_{kj}\;\alpha_{mj}$, which
proves equation (\ref{invrel1}) is satisfied, and the relations (\ref{det}) and (\ref{bc}) show
$ a d = d a$ for $q=-1$ which implies $\alpha_{ki}\;\alpha_{mj} = \alpha_{mj}\;\alpha_{ki}$ thus proving that
equation (\ref{invrel2}) is satisfied by the elements of an $SO_{q=-1}(3)$ matrix.

It is a little harder to show that equation (\ref{invrel3}) is satisfied by elements of $SO_{q=-1}(3)$ matrices.
However, if one writes out the indices of the equation, one finds that equation (\ref{invrel3}) implies that each
matrix element is equal to the $GL_{q=-1}(2)$-determinant of its minor. This fact is indeed satisfied by $SO_{q=-1}(3)$
matrices since $\mathrm{det} A = 1$ and ${A^{-1}}^T = A$, one can show that $A = Co(A)$ which itself means that every element is
equal to the determinant of its minor. Note that since $q = -1$, the cofactor of an element is always equal to the minor
without any alternation of signs. This type of determinant with no alternation of signs is also called a permanent.

Thus, we have found that the invariance quantum group of ACSA is the quantum group $SO_q(3)$ with $q=-1$.
Strictly speaking, the ACSA is a module of the $q$-deformed $SO(3)$ quantum algebra with $q = -1$. It is very
interesting to note that the invariance group of the angular momentum algebra is also $SO_q(3)$ but with $q=1$.

\section{Representations}

The Anticommutator Spin Algebra is defined by the relations (\ref{eqn:defrel1}-\ref{eqn:defrel3}). In order to find the representations
of this algebra we define the operators:
\bea
J_+ & = & J_1 + J_2 \\
J_- & = & J_1 - J_2 \\
J^2 & = & J_1^2 + J_2^2 + J_3^2
\eea
which obey the following relations:
\bea
\anti{J_+}{J_3} & = & J_3 \\
\anti{J_-}{J_3} & = & -J_3 \\
J_+^2 & = & J^2 - J_3^2 + J_3 \label{jp^2}\\
J_-^2 & = & J^2 - J_3^2 - J_3 \label{jm^2}
\eea
Furthermore, it can easily be shown that $J^2$ is central in the algebra, ie that it commutes with
all the elements of the algebra. For this reason, we can label the states in our representation with the
eigenvalues of $J^2$ and $J_3$:
\bea
J^2 \ket{\lambda, \mu} & = & \lambda \ket{\lambda, \mu} \\
J_3 \ket{\lambda, \mu} & = & \mu \ket{\lambda, \mu}
\eea
The action of $J_+$ and $J_-$ on the states such defined is easily shown to be:
\bea
J_+ \ket{\lambda, \mu} & = & f(\lambda, \mu) \ket{\lambda,- \mu + 1} \label{jplus}\\
J_- \ket{\lambda, \mu} & = & g(\lambda, \mu) \ket{\lambda,- \mu - 1} \label{jminus}
\eea
It is enough to look at the norm of the states $J_+ \ket{\lambda, \mu}$ and $J_- \ket{\lambda, \mu}$
to find $f(\lambda, \mu)$ and $g(\lambda, \mu)$. Thus:
\bea
\bra{\lambda, \mu} J_+^2 \ket{\lambda, \mu} & = & |f(\lambda, \mu)|^2 \\
\bra{\lambda, \mu} J^2 - J_3^2 + J_3 \ket{\lambda, \mu} & = & |f(\lambda, \mu)|^2 \\
\lambda - \mu^2 + \mu & = & |f(\lambda, \mu)|^2 \\
f(\lambda, \mu) & = & \sqrt{\lambda - \mu^2 + \mu}
\eea
and, similarly, $g(\lambda, \mu) = \sqrt{\lambda - \mu^2 - \mu}$. These coefficients must be
real due to the fact that $J_+$ and $J_-$ are hermitian operators. This constraint imposes the
following conditions on $\lambda$ and $\mu$:
\bea
\lambda - \mu^2 + \mu & \geq & 0 \\
\lambda - \mu^2 - \mu & \geq & 0
\eea
which can be satisfied by letting $\lambda = j(j+1)$ for some $j$ with:
\beq
j \geq \mu \geq -j. \label{muspec}
\eeq

Note that equation (\ref{jplus}) shows that the action of $J_+$ is composed of a reflection which changes sign of $\mu$,
the eigenvalue of $J_3$, followed by raising by one unit. Similarly, equation (\ref{jminus}) shows that $J_-$ reflects and lowers.
Thus the highest state $\mu = j$ is annihilated by $J_-$ and "lowered" by $J_+$. Applying $J_+$ or $J_-$ twice to any state
gives back the same state due to relations (\ref{jp^2}) and (\ref{jm^2}). Thus starting from the highest state we apply $J_-$ and $J_+$
alternately to get the spectrum:
\beq
j, -j+1, j-2, -j+3, ...
\eeq
This sequence ends so as to satisfy equation (\ref{muspec}) only for integer or half-integer $j$. For integer $j$,
it terminates, after an even number of steps, at $-j$ and visits every integer in between only once. For
half-integer $j=2k\pm\frac{1}{2}$ it ends at $j=\pm\frac{1}{2}$ having visited only half the states with $\mu$
half-odd integer between $j$ and $-j$. The rest of the states cannot be reached from these but are obtained by starting
from the $\mu= -j$ state and applying $J_-$ and $J_+$ alternately; starting with $J_-$.
\\
We now give a few examples:
\begin{itemize}
  \item
For $\mathbf{j=2}$ the states follow the sequence:
\[
\mathbf{\mu = 2, -1, 0, 1, -2}\quad.
\]
  \item
For $\mathbf{j=\frac{3}{2}}$ there exist two irreducible representations one with:
\[
\mathbf{\mu = \frac{3}{2}, -\frac{1}{2}}\quad,
\] and the other with:
\[
\mathbf{\mu = -\frac{3}{2}, \frac{1}{2}}\quad.
\]

  \item
For $\mathbf{j=\frac{5}{2}}$ the two representations are given by:
\[
\mathbf{\mu = \frac{5}{2}, -\frac{3}{2}, \frac{1}{2}}\quad,
\] and by:
\[
\mathbf{\mu = -\frac{5}{2}, \frac{3}{2}, -\frac{1}{2}}\quad.
\]

\end{itemize}

\section{Conclusions}

The Anticommutator Spin Algebra, which is a special Jordan algebra, described in this paper has many implications.
The first of these is the fact that this algebra is a consistent fermionic algebra which is not a superalgebra.
For possible physical applications
the right-hand side of the defining relations (\ref{eqn:defrel1}-\ref{eqn:defrel3}) must also be supplied
with an $\hbar$. In a superalgebra approach where the $J_i$ are regarded as odd operators, the $\hbar$
on the right-hand side should also be regarded as an operator anticommuting with the $J_i$. These models \cite{leites,batalin}
result from the quantization of the odd Poisson bracket. In our approach however, the concept of grading and therefore
an underlying Poisson bracket formalism does not exist. In particular, there is no Jacobi identity. Nevertheless,
the associative algebra we consider is consistent with quantum mechanics where physical observables correspond to
hermitian operators and their eigenvalues to possible results of physical measurement of these observables.
It is for this reason that ACSA suggests a new kind of statistics which, we believe, will be useful in physics.

The second implication is the important role of quantum groups in mathematical physics. As we have shown
in this paper, the invariance group of ACSA turns out to be a quantum group.
Given the fact that ACSA is very similar to normal spin algebra and that the invariance group of spin algebra
plays an important role in physics, the invariance quantum group of ACSA, $SO_{q=-1}(3)$, becomes a prime example of how
central quantum groups have become in mathematical physics. It is also interesting to note that more algebras like
ACSA can be constructed where the commutators of the original Lie algebra are turned into anticommutators and that
such algebras might also have invariance quantum groups that is the same as the invariance group of the original Lie
algebra with $q=-1$. This possibility is open to investigation in a more general framework.

\end{document}